\documentclass[aps,pra,twocolumn,showpacs,preprintnumbers,amsmath,amssymb]{revtex4-2}
\usepackage{soul}
\usepackage{amsmath}
\usepackage{esvect}

\usepackage{natbib}
\usepackage{graphicx}
\usepackage{color}
\usepackage{tabularx}
\usepackage{multirow}

\usepackage{wasysym}

\usepackage{amssymb}

\usepackage[utf8]{inputenc}

\usepackage{float}

\usepackage{subfigure}

\usepackage{float}

\usepackage{hyperref}

\usepackage{cancel}

\usepackage{xcolor}

\usepackage{setspace}
\usepackage{soul}
\usepackage{wrapfig}

\definecolor{burntorange}{rgb}{0.8, 0.33, 0.0}

\begin{document}

\title{Frequency locking: a distinctive feature of the coherent population trapping\\
and the stationarity effect}

\author{E.\,A.~Tsygankov$^{1}$}
\email[]{tsygankov.e.a@yandex.ru}
\author{D.\,S.~Chuchelov$^{1}$}
\author{M.\,I.~Vaskovskaya$^{1}$}
\author{V.\,V.~Vassiliev$^{1}$}
\author{S.\,A.~Zibrov$^{1}$}
\author{V.\,L.~Velichansky$^{1}$}
\affiliation{1. P.\,N. Lebedev Physical Institute of the Russian Academy of Sciences,\\
Leninsky Prospect 53, Moscow, 119991 Russia}

\newcommand{\red}{\textcolor{red}}
\newcommand{\blue}{\textcolor{blue}}

\begin{abstract}

We~study the case where phase modulation of~the harmonic signal is~used to~obtain the error signal for the frequency stabilization to~a~reference atomic transition. High-frequency modulation, or~analog of~the Pound-Drever-Hall regime, is~considered. We demonstrate that for coherent population trapping, the maximal error-signal slope retains at~a~certain level with growth in~the modulation frequency, while for other types of~resonances it~drops steadily. The investigation of~the low-frequency modulation regime reveals the stationarity effect. We~show that in~this case, the maximal steepness of~the error signal does not depend on~the modulation frequency and is~reached at~a~fixed value of~the frequency deviation.
\end{abstract}

\maketitle

\section{Introduction}

The basic idea behind the microwave~\cite{vanier2015quantum,riehle2006frequency} or~optical~\cite{ludlow2015optical} frequency standard is~to~compare the local oscillator frequency with that of~a~reference atomic transition. Among the former, the most widespread and used are clocks based on~microwave transition between magneto-insensitive sublevels (in~the linear approximation) in~the ground state $\textbf{nS}_{1/2}$ of~alkali-metal atoms. The classic example is~the rubidium atomic frequency standards, which utilize the double radio-optical resonance~\cite{riley2019history}. The pumping light source of~$^{87}$Rb atoms enclosed in~a~glass-blowing cell is~a~discharge lamp, and the reference transition is~interrogated by~an~rf~field from the microwave cavity. Designed and commercialized in~the $1960$s, they are now the workhorse of~the telecommunications industry and satellites of~global navigation systems~\cite{riley2019history,batori2021gnss}.

The all-optical effect of~coherent population trapping (CPT), discovered in~the $1970$s~\cite{Alzetta1976,PhysRevA.14.1498,1976NCimL..17..333A,Gray:78,arimondo1996v}, has led to~the development of~chip-scale atomic clocks in~the $2000$s and their commercialization in~the early $2010$s~\cite{cash2018microsemi}. The pumping light produced by~the current-modulated vertical-cavity surface-emitting diode laser~\cite{affolderbach2000nonlinear}, the microfabricated glass cell with alkali-metal atoms~\cite{knappe2007chip} and no~need to~use the microwave cavity~\cite{vanier2005atomic} are the reasons why the CPT-based atomic clocks became unrivaled in~size, weight, and power consumption among the frequency standards~\cite{9316270}. This feature determined the use of~such devices in~unmanned vehicles, underwater sensor systems, small satellites, military applications, etc.~\cite{Microsemi,JRC125394}.

The long-term frequency stability of~rubidium and chip-scale atomic clocks is~hard to~improve. Nowadays, there are emerging compact optical clocks since the stabilized laser frequency can be~moved down to~the consumer range by~means of~frequency combs based on~high-quality microresonators~\cite{newman2019architecture}.
Such a~frequency standards are also based on~glass cells with alkali-metal atoms. Two notable examples are clocks utilizing the two-photon transition in~$^{87}$Rb atoms~\cite{PhysRevApplied.9.014019} and those that use sub-Doppler resonance in~the alkali-metal atoms D$_1$ line induced by~counter-propagating bichromatic laser waves~\cite{gusching2023cs}. Both approaches are promising to~provide the frequency stability at~the level of~$10^{-13}$ at~$1$~s (ADEV).

For all the clocks mentioned above, the reference resonance is~an~even function of~the detuning, and therefore it~cannot be~directly used to~stabilize the local oscillator frequency. One common approach to~obtain the error signal with a~dispersive shape is~to~use the lock-in technique based on~the phase modulation of~the interrogating field's frequency, \hbox{$s(t)=\cos{\left(\omega_0t+m\sin{\omega_mt}\right)}$}. Here $m$ is~the phase modulation index, $\omega_m$ is~the modulation frequency. As~a~result the light absorption oscillates at~frequencies $n\omega_m$, $n>0$, among which the first harmonic $n=1$ is~most often used. Amplitude of~the corresponding oscillations can be~represented as~the sum of~the in-phase ($\propto\cos{\omega_mt}$) and quadrature ($\propto\sin{\omega_mt}$) signals. They can be obtained from the absorption signal by the synchronous detection technique. In~a~general case a~mixture like $A_{\text{In-ph}}\cos{\alpha}-A_{\text{Q}}\sin{\alpha}$, where $A_{\text{In-ph}}$ and $A_{\text{Q}}$ are amplitudes of~the signals, should be~used to~maximize slope of~the error signal, where $\alpha$ is~the synchronous detection phase.

It~is~convenient to~treat $s(t)$ in~the spectral interpretation using the Jacobi-Anger expansion, \hbox{$e^{-im\sin{\omega_m}t}=\sum^{\infty}_{k=-\infty}J_k(m)e^{-ik\omega_mt}$}, where $J_k(\cdot)$ is~the Bessel function of~the first kind of~the index $k$. In~this case one can demonstrate, that non-oscillating part of~the light absorption is~a~group of~Lorentzians spaced by~$\omega_m$, i.e.,~it acquires a~multipeak structure. As~we~have recently shown for the case of~the CPT resonance~\cite{PhysRevA.109.053703}, this feature leads to~the frequency pulling of~the error signal under an asymmetry. However, this effect can be~suppressed in~a~high-frequency modulation,~or analog of~the Pound-Drever-Hall regime~\cite{drever1983laser}, where the peaks are well resolved. The question remains in~the degree~of the error-signal slope reduction compared to~the maximal possible value, which is~achieved when the peaks are not resolved.

\vbox{In~this work we~show that the maximal slope of~the error signal steadily drops when the modulation frequency is~increased beyond the width of~the optical transition in~two-level system and the width of double-radio optical resonance. The CPT effect, on the contrary, turns out to~be~unique: the steepness retains at~a~certain level with $\omega_m$ growth in~the high-frequency regime. We~also demonstrate that for all systems of levels mentioned above there is~the stationarity effect: the maximal error-signal slope is~constant when $\omega_m$ is~much smaller than the interrogated transition width.}

\begin{figure}[ht] 
  \center
  \includegraphics[width=0.45\textwidth]{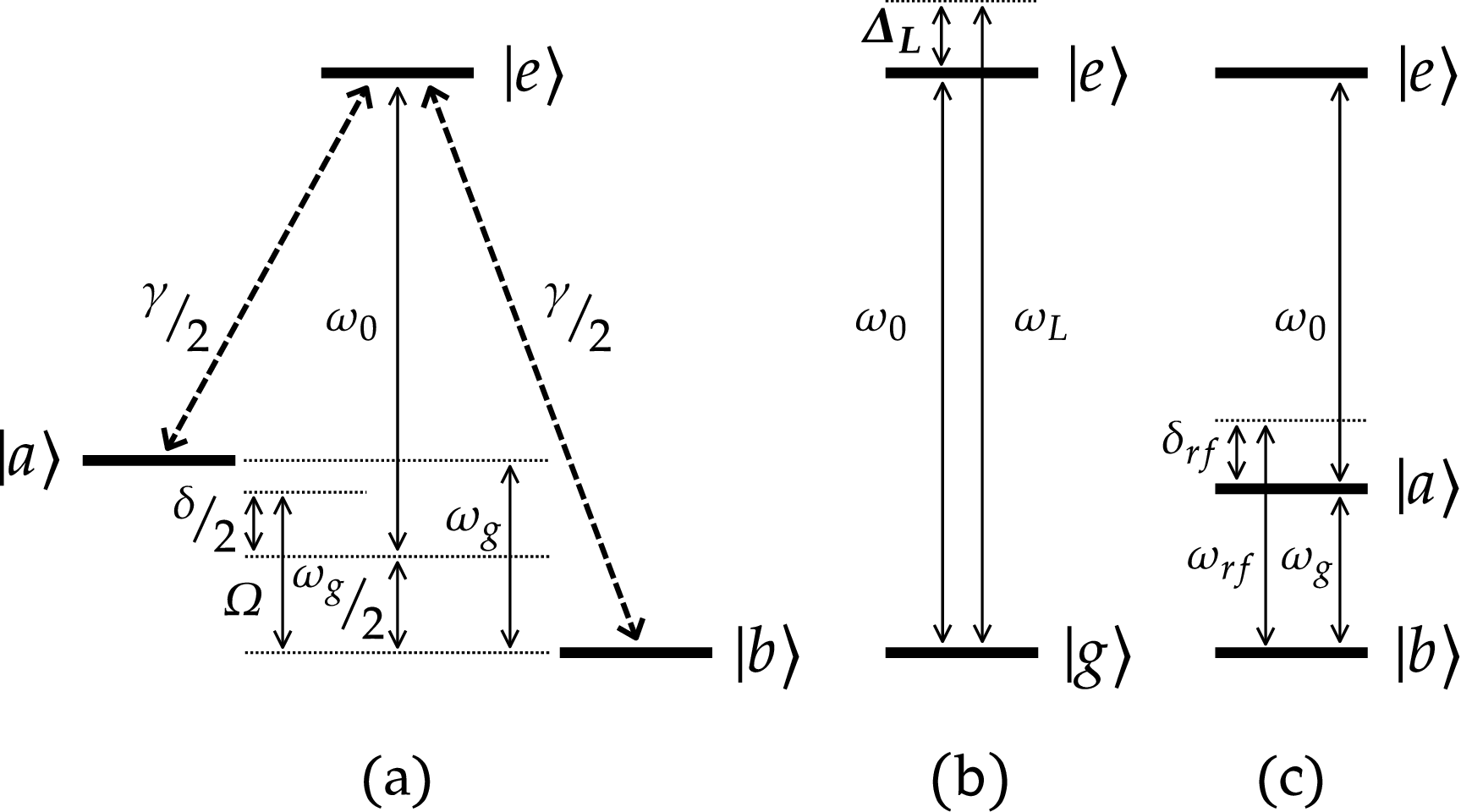}
  \caption{Systems of levels under consideration.}
  \label{Schemes}
\end{figure}

We begin from the CPT effect and consider the $\Lambda$-system of~levels: the excited-state level is~$|e\rangle$, ground-state levels $|a\rangle$ and $|b\rangle$ are spaced by~the frequency interval $\omega_g$; see Fig.~\ref{Schemes}a. The following bichromatic optical field
\begin{equation}
\mathcal{E}(t)=-\dfrac{\mathcal{E}_0}{2}\left\{e^{-i\left[(\omega_0-\Omega)t-\varphi(t)\right]}-e^{-i\left[(\omega_0+\Omega)t+\varphi(t)\right]}\right\}+\text{c.\,c.},
\end{equation}

\noindent couples $|a\rangle$ and $|b\rangle$ with $|e\rangle$. The low-frequency component induces transitions only between levels $|a\rangle$ and $|e\rangle$, while the high-frequency component works on the opposite shoulder of the $\Lambda$-scheme. The frequency $\omega_0$ is~taken to~be~equal to~half-sum of~the optical transitions and $\omega_0\gg\omega_g$. The dipole moments of~both transitions are assumed to~be~equal and real. The term $\varphi(t)=m\sin{\omega_m}t$ accounts for the modulation required for producing in-phase and quadrature signals. The frequency $\Omega$ is~close to~$\omega_g/2$, and the difference gives the two-photon detuning $\delta=2\Omega-\omega_g$.

The initial equations for the density matrix elements can be~treated by~adiabatically eliminating the excited state, using the rotating wave approximation, and assuming the low saturation regime. Here we~cite the simplified equations obtained in our previous work~\cite{PhysRevA.109.053703}:
\begin{equation}
\rho_{ee}=2\dfrac{V^2}{\gamma\Gamma}\left\{1-2\text{Re}\left[e^{2i\varphi(t)}\bar{\rho}_{ab}(t)\right]\right\},
\end{equation}
\begin{equation}
\left(i\frac{\partial}{\partial t}+\delta+i\tilde{\Gamma}_g\right)\bar{\rho}_{ab}=i\dfrac{V^2}{\Gamma}e^{-2i\varphi(t)},
\label{rhoab2}
\end{equation}

\noindent where $\rho_{ee}$ is~the excited-state population and $\bar{\rho}_{ab}$ is~the slow amplitude of~the non-diagonal element of~the density matrix between levels of~$\Lambda$-system's ground state, $\bar{\rho}_{ab}=\rho_{ab}e^{-2i\Omega t}$. $V=d\mathcal{E}_0/2\hbar$ is~the Rabi frequency, $\gamma$ is~the spontaneous decay rate of~the excited-state population, $\Gamma$ is~the relaxation rate of~the optical coherences, which describes homogeneous broadening of~the optical line. The relaxation parameter $\tilde{\Gamma}_g=\Gamma_g+2V^2/\Gamma$ accounts for the power broadening of~the CPT resonance by~the optical field. The hierarchy of~parameters is~the following:
\begin{equation}
\delta,\,\tilde{\Gamma}_g,\,\omega_m,\,V^2/\Gamma\ll\gamma\ll\Gamma\ll\Omega,\,\omega_g\ll\omega_0.
\end{equation}

\begin{figure*}[ht]

\centerline{
\subfigure[]{\label{CPT-SlopeWithInset}
\includegraphics[width=0.5\textwidth]{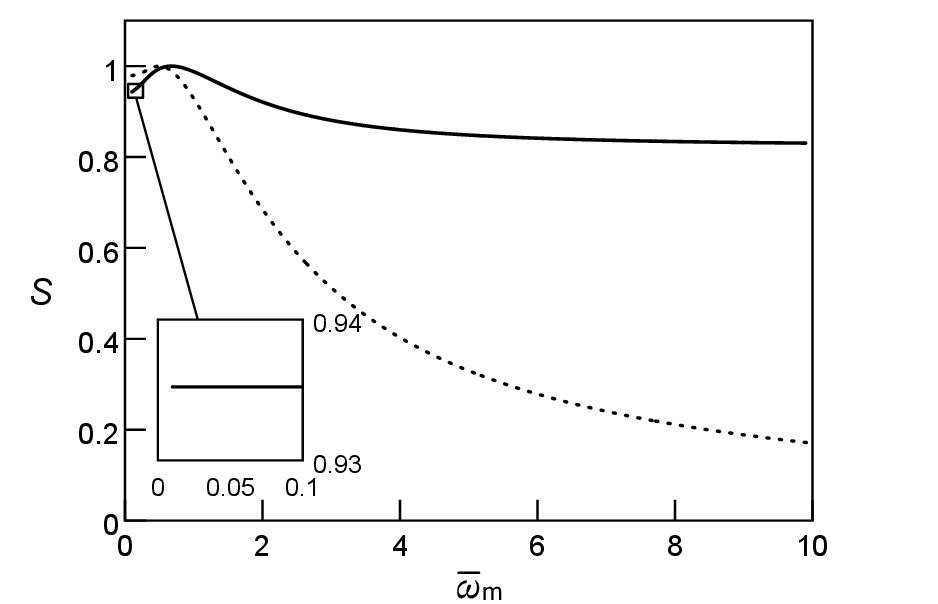}}
\hfill
\subfigure[]{\label{CPT-IndexAndPhase}\includegraphics[width=0.5\textwidth]{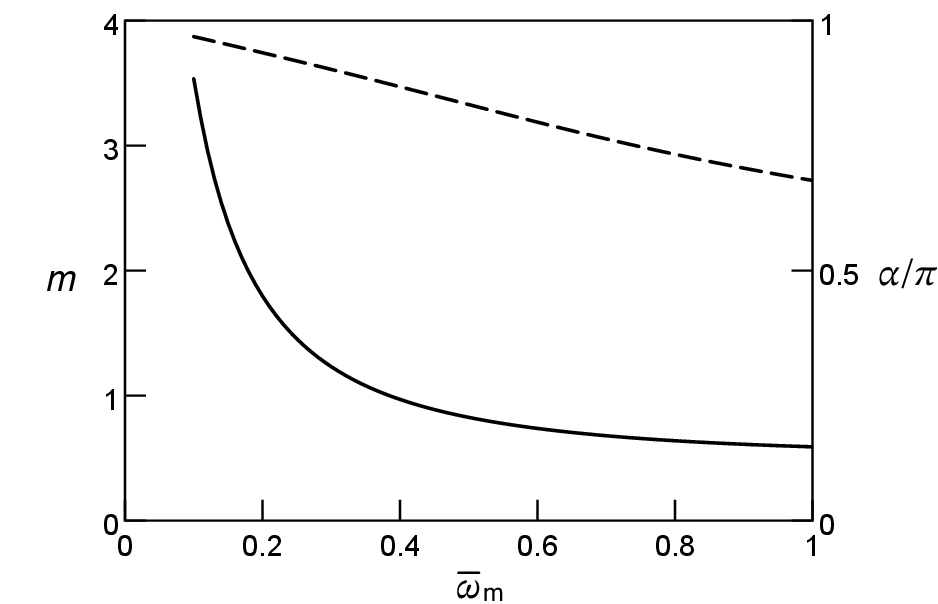}}}

\caption{(a) Solid curve: dependence of~the maximal error-signal slope on~the ratio of~the modulation frequency $\omega_m$ to~the ground-state relaxation rate $\tilde{\Gamma}_g$ for the CPT resonance. The maximum value is~normalized to~be~unity. The inset demonstrates the low-frequency region, the curve is~plotted via~Eq.~\eqref{CPT-InPhase}. Dashed curve: the same but for the two-level system, where $\bar{\omega}_m=\omega_m/(\gamma/2)$. (b) The corresponding dependence of~the modulation index $m$ (solid curve) and the synchronous detection phase $\alpha$ (in~units of~$\pi$, dashed curve) providing the maximal error-signal slope for the CPT resonance.}
\label{CPTFigure}

\end{figure*}

By using the Jacobi-Anger expansion and the Fourier one for $\rho_{ee}$ and $\bar{\rho}_{ab}$ over $\omega_m$, we~get the following expression determining response of~the system at~the frequency~$\omega_m$:
\begin{equation}
\begin{gathered}
A^{\text{CPT}}_1=-4\dfrac{V^2}{\gamma\Gamma}\text{Re}\bigg\{i\sum^{\infty}_{k=-\infty}J_k(2m)\left[J_{k-1}(2m)e^{-i\omega_mt}\right.\\
\left.+J_{k+1}(2m)e^{i\omega_mt}\right]\dfrac{V^2/\Gamma}{\delta+k\omega_m+i\tilde{\Gamma}_g}\bigg\}.
\label{A1CPT}
\end{gathered}
\end{equation}

The amplitudes of~in-phase and quadrature signals in~the experiment are obtained by the synchronous detection technique:
\begin{equation}
\langle\rho_{ee}\cos{(\omega_mt+\alpha)}\rangle,
\end{equation}

\noindent where $\langle\dots\rangle$ means averaging over period of~the frequency $\omega_m$. In~the case of~numerical results, presented in~Figs.~\ref{CPT-SlopeWithInset},~\ref{CPT-IndexAndPhase}, we~have investigated the range $\alpha\in(0,\,\pi)$. The absolute value of~steepness was obtained by~linearization of~the in-phase and quadrature signals amplitudes over $\delta$. The maximal slope is~reached at~$\omega_m/\tilde{\Gamma}_g\simeq0.764$, $m\simeq0.652$ and $\alpha\simeq3\pi/4$, which is~in~accordance with work~\cite{Yudin2017}.

As~we~will see further, in~comparison to the case of~the two-level system and the double radio-optical resonance, the expression~\eqref{A1CPT} does not contain the term like $1/(\omega_m+i\tilde{\Gamma}_g)$ before the sum. This results in~the constant level of the error-signal steepness at~$\omega_m\gg\tilde{\Gamma}_g$; see Fig.~\ref{CPT-SlopeWithInset}. In~the high-frequency modulation regime, the slope is~given by~the central dispersive curve of~the quadrature signal, which is~proportional to~the product $J_0(2m)J_1(2m)$ and does not depend on~$\omega_m$; see~\cite{PhysRevA.109.053703}. The maximizing value of~$m$ is~$\simeq0.54$. Dependencies, presented in~Figs.~\ref{CPT-SlopeWithInset},~\ref{CPT-IndexAndPhase}, were numerically obtained by~accounting terms from $k=-100$ to~$k=100$ in~the sums.

The maximizing value of~the modulation index $m$ grows when the frequency $\omega_m$ is~decreased beyond $\tilde{\Gamma}_g$; see Fig.~\ref{CPT-IndexAndPhase}. This feature stems from the structure of~the in-phase signal:
\begin{equation}
\begin{gathered}
A^{\text{CPT}}_{\text{In-ph}}\propto\sum^{\infty}_{k=-\infty}J_k(2m)\dfrac{J_{k-1}(2m)+J_{k+1}(2m)}{(\bar{\delta}+k\bar{\omega}_m)^2+1}\\
\equiv-4\bar{\delta}\dfrac{\bar{\omega}_m}{m}\sum^{\infty}_{k=1}\dfrac{\left[kJ_k(2m)\right]^2}{\left[(\bar{\delta}+k\bar{\omega}_m)^2+1\right]\left[(\bar{\delta}-k\bar{\omega}_m)^2+1\right]}\\
\underset{\bar{\delta}\ll1}{\simeq}
-4\bar{\delta}\dfrac{\bar{\omega}_m}{m}\sum^{\infty}_{k=1}\left[\dfrac{kJ_k(2m)}{\left(k\bar{\omega}_m\right)^2+1}\right]^2,
\end{gathered}
\label{CPT-InPhase}
\end{equation}

\noindent where $\bar{\delta}=\delta/\tilde{\Gamma}_g$, $\bar{\omega}_m=\omega_m/\tilde{\Gamma}_g$. The first line of~this formula demonstrates that dispersive shape of~the signal stems from pairs of~Lorentzians symmetrically located with respect to~the point $\bar{\delta}=0$ but having opposite signs. The slope of~each pair is~maximized when they are shifted by $\simeq0.58\tilde{\Gamma}_g$ from the point $\bar{\delta}=0$. Therefore, with decrease in~$\bar{\omega}_m$, the slope is~determined by~pairs of~Lorentzian given by~greater $|k|$; see the expression in~the third line of~Eq.~\ref{CPT-InPhase}. In~its turn, for $k\gg1$, the corresponding Bessel functions are maximized at~$m\propto k$. Therefore, the value of~$m$ maximizing the steepness grows as~$1/\bar{\omega}_m$ under decreasing of~$\bar{\omega}_m$ at~$\bar{\omega}_m\ll1$. Fig.~\ref{CPT-IndexAndPhase} demonstrates that the corresponding dependence of~$m$ on~$\bar{\omega}_m$ is~the hyperbola. The numerical calculation reveals that the maximal error-signal slope provided by~the in-phase signal is~constant at~$\bar{\omega}_m\ll1$. In~the time representation this means that the frequency deviation does not change.

We~continue with the two-level system considering the case of~the optical transition between the single ground level $|g\rangle$ and the excited state $|e\rangle$ (see Fig.~\ref{Schemes}b) induced by~the phase-modulated field:
\begin{equation}
\mathcal{E}(t)=\mathcal{E}_0e^{-i\left[\omega_Lt+\varphi(t)\right]}/2+\text{c.\,c.}
\end{equation}

The equations for the density matrix elements under the rotating wave approximation and the low saturation regime are the following:
\begin{equation}
i\dfrac{\partial}{\partial t}\rho_{ee}=V\left(e^{-i\varphi(t)}\rho_{ge}-\text{c.\,c.}\right)-i\gamma\rho_{ee},
\end{equation}
\begin{equation}
\left(i\dfrac{\partial}{\partial t}+\Delta_L+i\gamma/2\right)\rho_{eg}=Ve^{-i\varphi(t)}.
\end{equation}

Here $V=d\mathcal{E}_0/2\hbar$ is~the Rabi frequency, $\gamma$ is~the natural width of~the excited state, and $\Delta_L=\omega_L-\omega_0$ is~the frequency detuning. The spacing between the excited and ground states is~$\omega_0$. This equations are valid for $\gamma,\Delta_L,V\ll\omega_0$.

The amplitudes of~the in-phase and quadrature signals are determined by~the real and imaginary parts of~the following expression:
\begin{equation}
\begin{gathered}
A^{\text{Two-level}}_1=\dfrac{V^2}{\omega_m+i\gamma}\\
\sum^{\infty}_{k=-\infty}J_k(m)\left[\dfrac{J_{k+1}(m)}{\Delta_L+k\omega_m-i\gamma/2}-\dfrac{J_{k-1}(m)}{\Delta_L+k\omega_m+i\gamma/2}\right].
\label{A1Two-Level}
\end{gathered}
\end{equation}

As~far as~the term before the sum has $\omega_m$~in the denominator, the maximal steepness of~the error signal falls in~the high-frequency modulation regime. We~demonstrate this explicitly at~strong inequality $\omega_m\gg\gamma/2$, where the error-signal steepness is~determined by~the central dispersive curve, which amplitude reads as~\begin{equation}
A_{\text{C}}=2J_0(m)J_1(m)\dfrac{S}{\bar{\omega}_m}\dfrac{\bar{\Delta}_L}{\bar{\Delta}^2_L+1},
\label{central}
\end{equation}

\noindent where $\bar{\Delta}_L=\Delta_L/(\gamma/2)$, $\bar{\omega}_m=\omega_m/(\gamma/2)$, \hbox{$S=\left[V/(\gamma/2)\right]^2$}. Eq.~\eqref{central} shows linear decrease of~$A_{\text{C}}$ with $\bar{\omega}_m$.

The stationarity effect also takes place. We~get for $\bar{\omega}_m\ll1$, $\bar{\Delta}_L\ll1$ that the biggest term of~the in-phase signal amplitude is~given by~\begin{equation}
A^{\text{Two-level}}_{\text{In-ph}}=-4\bar{\Delta}_L
S
\dfrac{\bar{\omega}_m}{m}\sum^{\infty}_{k=1}
\left[\dfrac{kJ_k(m)}{\left(k\bar{\omega}_m\right)^2+1}\right]^2,
\end{equation}
having the sum of the same structure as for the CPT resonance.

Further, we consider the double radio-optical resonance in~the three-level system with the excited state $|e\rangle$ and two ground states $|a\rangle$, $|b\rangle$ spaced by~the interval $\omega_g$; see Fig.~\ref{Schemes}c. The optical field \hbox{$\mathcal{E}(t)=\mathcal{E}_0e^{-i\omega_0t}/2+\text{c.\,c.}$} induces only optical transitions $|a\rangle\rightarrow|e\rangle$ in~the exact resonance, while the rf~field $\mathcal{B}_{rf}(t)=\mathcal{B}_{rf}e^{-i\left[\omega_{rf}t+\varphi(t)\right]}/2+\text{c.\,c.}$ induces transitions between levels of the ground state. The spontaneous decay of~the upper level equally populates $|a\rangle$ and $|b\rangle$.

By~using the same approximations as~earlier, we~arrive at~the following equations:
\begin{equation}
\rho_{ee}=2\dfrac{V^2}{\gamma\Gamma}\rho_{aa},
\end{equation}
\begin{equation}
\begin{gathered}
i\dfrac{\partial}{\partial t}\rho_{aa}=V_{rf}\left[e^{-i\varphi(t)}\rho_{ba}-\text{c.\,c.}\right]\\
-i\dfrac{V^2}{\Gamma}\rho_{aa}-i\Gamma_g\left(\rho_{aa}-1/2\right),
\end{gathered}
\end{equation}
\begin{equation}
\left(i\dfrac{\partial}{\partial t}+\delta_{rf}+i\Gamma_g+i\dfrac{V^2}{\Gamma}\right)\rho_{ab}=V_{rf}e^{-i\varphi(t)}\left(1-2\rho_{aa}\right).
\end{equation}

The system above is~valid for the case of~small width of~optical transitions compared to~the ground-state interval, $\Gamma\ll\omega_g$. The rf~field detuning and the Rabi frequency are $\delta_{rf}=\omega_{rf}-\omega_g$ and $V_{rf}=\mu\mathcal{B}_{rf}/2\hbar$, respectively. The constant $\Gamma_g$ describes the relaxation of~the ground-state elements. The analytical solution can be~obtained at~$\rho_{aa}\ll1$ when the optical pumping is~sufficient, $V^2/\Gamma\gg\Gamma_g$ and the rf~field is~weak, $V_{rf}\ll V^2/\Gamma$.

Amplitudes of~the in-phase and quadrature signals are determined by~real and imaginary parts of~the following expression:
\begin{equation}
\begin{gathered}
A^{\text{DR}}_1=\dfrac{V^2_{rf}}{\omega_m+iV^2/\Gamma}\sum^{\infty}_{k=-\infty}J_k(m)\left(\dfrac{J_{k+1}(m)}{\tilde{\delta}_{rf}+k\omega_m-iV^2/\Gamma}\right.\\
\left.-\dfrac{J_{k-1}(m)}{\tilde{\delta}_{rf}+k\omega_m+iV^2/\Gamma}\right).
\end{gathered}
\end{equation}

We~have the same structure of~$A^{\text{DR}}_1$ as~for the two-level system. This is~not surprising, since under the chosen assumptions, the equations for the two cases are almost identical. Therefore we~conclude that the maximal steepness of~the error signal also drops with an~increase in~$\omega_m$, and the stationarity effect takes place for $\omega_m\ll V^2/\Gamma$.

To~summarize. As we~have demonstrated, the maximal error-signal slope does not fall only for the CPT effect in~the high-frequency modulation regime. This feature has the following explanation. The bichromatic optical field interrogates the microwave transition. However, the width of~the spectrum produced by~the phase modulation should be~compared with the optical transitions width. Therefore, a~decline in~the maximal slope should begin when $\omega_m$ exceeds $\gamma$. Thus, the high-frequency modulation regime is~very suitable for chip-scale atomic clocks to~suppress the frequency pulling effect and $1/f$ noise. The pros and cons of~high modulation frequencies for other types of frequency standards must be~carefully estimated.

Considering the stationarity effect, it~can be~useful in~cases where the modulation frequency cannot be~increased greater than the reference transition width. This, for example, is~the field of~devices where the direct modulation of~the diode laser current is~required for the stabilization of its frequency. Therefore, one could retain $\omega_m$ in~a~region far smaller than the optical transition width without significant loss in~the maximal error-signal steepness while reducing $1/f$ noise and avoiding the frequency pulling due to~an~asymmetric multipeak structure.


For completeness, we~remind here that the maximal steepness also falls with $\omega_m$ in~the high-frequency regime when the laser frequency is~stabilized to~the transmission peak of~an~interferometer (an~infinite free spectral range is~assumed here). But since the Pound-Drever-Hall technique utilizes the reflected signal, for which the steepness of~the central dispersive curve grows as~$\bar{\omega}^2_m/(1+\bar{\omega}^2_m)$ with $\omega_m$ (see, for example, formulae $9.19$ in~\cite{riehle2006frequency}), there are no~issues with the high-frequency modulation regime. We~note that the steepness of~the central dispersive curve does not depend on~$\omega_m$ only in~the case of~the CPT resonance among the considered systems.

\end{document}